\documentstyle[12pt]{article}
\newcommand{\be}{\begin{equation}}
\newcommand{\ee}{\end{equation}}
\newcommand{\er}{\end{eqnarray}}
\newcommand{\br}{\begin{eqnarray}}
\newcommand{\dslash}{\partial\!\!\!/}
\newcommand{\aslash}{A\!\!\!/}
\newcommand{\bslash}{B\!\!\!/}
\newcommand{\kslash}{\kappa\!\!\!/}
\newcommand{\fslash}{f\!\!\!/}
\newcommand{\Dslash}{D\!\!\!\!/}
\begin{document}
\thispagestyle{empty}
hep-th/9709105\\
$\phantom{x}$\vskip 0.618cm\par
{\huge \begin{center}Bosonisation and Soldering of Dual Symmetries in
Two and Three Dimensions
\end{center}}\par

\begin{center}
$\phantom{X}$\\
{\Large R.Banerjee\footnote{On leave
of absence from S.N.Bose National Centre for Basic
Sciences, Calcutta, India. e-mail:rabin@if.ufrj.br} and C.Wotzasek}\\[3ex]
{\em Instituto de F\'\i sica\\
Universidade Federal do Rio de Janeiro\\
21945, Rio de Janeiro, Brazil\\}
\end{center}\par
\begin{abstract}

\noindent We develop a technique that solders the dual aspects of some
symmetry following from the bosonisation of two distinct fermionic
models, thereby leading to new results which cannot be otherwise
obtained. Exploiting this technique, the two dimensional chiral
determinants with opposite chirality are soldered to reproduce
either the
usual gauge invariant expression leading to the Schwinger model or,
alternatively, the Thirring model. Likewise, two apparently
independent three dimensional massive Thirring models with same
coupling but opposite mass signatures, in the long wavelegth limit,
combine by the process of bosonisation and soldering
to yield an effective massive Maxwell theory. 
The current bosonisation formulas are given, both in
the original independent formulation as well as the effective theory, and
shown to yield consistent results for the correlation functions.
Similar features
also hold for quantum electrodynamics in three dimensions.
\end{abstract}

Keywords: Bosonisation; Soldering; Dual symmetry

PACS number: 11.15 
\vfill
\newpage

\section{Introduction}
Bosonisation is a powerful technique that  maps a
fermionic theory into its bosonic counterpart. 
It was initially developed and fully
explored in the context of two dimensions\cite{AAR}. More recently, it has been
extended to higher dimensions\cite{M,C,RB,RB1}.
The importance of bosonisation lies in
the fact that it includes quantum effects already at the classical level. 
Consequently, different aspects and manifestations of quantum phenomena
may be investigated directly, that would otherwise be highly nontrivial
in the fermionic language. Examples of such applications are the 
computation of the current algebra\cite{RB} and the study of screening or
confinement in gauge theories\cite{AB}.

This paper is devoted to analyse certain features and applications of
bosonisation which, as far as we are aware, are unexplored even in two
dimensions. The question we pose is the following: given two independent
fermionic models which can be bosonised separately, under what 
circumstances is it possible to represent them by one single effective theory? 
The answer lies in the symmetries of the problem. Two  distinct models
displaying dual aspects of some symmetry can be combined by the
simultaneous implementation of bosonisation and soldering to yield a
completely new theory. This is irrespective of dimensional considerations.
The technique of soldering essentially comprises in lifting the gauging
of a global symmetry to its local version and exploits certain concepts
introduced in a different context by Stone\cite{S} and one of us\cite{W}.
The analysis is intrinsically
quantal without having any classical analogue. This is easily explained
by the observation that a simple addition of two independent classical
lagrangeans is a trivial operation without leading to anything meaningful
or significant.

The basic notions and ideas are first introduced in the context of two
dimensions where bosonisation is known to yield exact results. The 
starting point is to take two distinct chiral lagrangeans with opposite
chirality. Using their bosonised expressions, the soldering mechanism
fuses, in a precise way, the left and right chiralities. This leads to
a general lagrangean in which the chiral symmetry no longer exists, but
it contains  two extra parameters manifesting the bosonisation ambiguities.
It is shown that different  parametrisations lead to different
models. In particular, the gauge invariant Schwinger model and Thirring
model are reproduced. As a byproduct, the importance of Bose symmetry
is realised and some interesting consequences regarding the arbitrary
parametrisation in the chiral Schwinger model are charted.

Whereas the two dimensional analysis lays the foundations, the subsequent
three dimensional discussion illuminates the full power and utility of
the present approach. While the bosonisation in these dimensions is not
exact, nevertheless, for massive fermionic models in the large mass or,
equivalently, the long wavelength limit, well defined local expressions
are known to exist\cite{C,RB}. Interestingly, these expressions exhibit a self
or an anti self dual symmetry that is dictated by the signature of the fermion
mass. Clearly, therefore, this symmetry simulates the dual aspects of
the left and right chiral symmetry in the two dimensional example,
thereby providing a novel testing ground for our ideas. Indeed, two distinct
massive Thirring models with opposite mass signatures,
are soldered to yield a massive Maxwell theory. This result
is vindicated by a direct comparison of the current correlation functions
obtained before and after the soldering process. As another instructive
application, the fusion of two models describing quantum electrodynamics
in three dimensions is considered. Results similar to the corresponding
analysis for the massive Thirring models are obtained.

We conclude by discussing future prospects and possibilities of extending
this analysis in different directions.

\bigskip

\section{The two dimensional example}

\bigskip

In this section we develop the ideas in the context of  two dimensions.
Consider, in particular, the following lagrangeans with opposite chiralities, 
\begin{eqnarray}
{\cal L}_+&=&\bar\psi(i \dslash + e \aslash P_+)\psi\nonumber\\
{\cal L}_-&=&\bar\psi(i \dslash + e \aslash P_-)\psi
\label{10}
\end{eqnarray}
where $P_\pm$ are the projection operators,
\begin{equation}
P_\pm=\frac{1 \pm \gamma_5}{2}
\label{20}
\end{equation}
It is well known that the computation of the fermion determinant, which
effectively yields the bosonised expressions, is plagued by regularisation
ambiguities since chiral gauge symmetry cannot be preserved\cite{RJ}. Indeed an
explicit one loop calculation following Schwinger's point splitting method
\cite{RB2} yields,
\begin{eqnarray}
\label{30}
W_+[\varphi] &=&-i \log \det (i\dslash+e\aslash_+)= 
{1\over{4\pi}}\int d^2x\,\left(\partial_+
\varphi\partial_-\varphi +2 \, e\,A_+\partial_-\varphi + a\, 
e^2\, A_+ A_-\right)\nonumber\\
W_-[\rho]&=&-i \log \det (i\dslash+e\aslash_-)= 
{1\over{4\pi}}\int d^2x\,\left(\partial_+\rho\partial_-
\rho +2 \,e\, A_-\partial_+\rho
+ b\, e^2\, A_+ A_-\right)
\end{eqnarray}
where the light cone metric has been invoked for convenience,
\begin{equation}
\label{35}
A_\pm = {1\over\sqrt 2}(A_0\pm A_1)=A^\mp \;\;\; ;\;\;\; \partial_\pm=
{1\over\sqrt 2}(\partial_0\pm \partial_1)=\partial^\mp
\end{equation}
Note that the regularisation or bosonisation ambiguity is manifested through
the arbitrary parameters $a$ and $b$. The latter ambiguity is particularly
transparent since by using the normal bosonisation dictionary
$\bar\psi i\dslash\psi \rightarrow \partial_+\varphi\partial_-\varphi$
and $\bar\psi\gamma_\mu\psi\rightarrow{1\over\sqrt \pi}\epsilon_{\mu\nu}
\partial^\nu\varphi$ (which holds only for a gauge invariant theory),
the above expressions with $a=b=0$ are easily reproduced from (\ref{10}).

It is crucial to observe that different scalar fields $\phi$ and $\rho$
have been used in the bosonised forms to emphasize the fact that the
fermionic fields occurring in the chiral components are uncorrelated.
It is the soldering process which will abstract a meaningful combination
of these components\cite{ABW}. This process essentially consists in lifting the
gauging of a global symmetry to its local version. Consider, therefore,
the gauging of the following global symmetry,
\begin{eqnarray}
\label{40}
\delta \varphi &=& \delta\rho=\alpha\nonumber\\
\delta A_{\pm}&=& 0
\end{eqnarray}

\noindent The variations in the effective actions  (\ref{30}) are found to be,

\begin{eqnarray}
\label{50}
\delta W_+[\varphi] &=& \int d^2x\, \partial_-\alpha \;J_+
(\varphi)\nonumber\\
\delta W_-[\rho]&=& \int d^2x\, \partial_+\alpha \;J_-(\rho)
\end{eqnarray}

\noindent  where the currents are defined as,

\begin{equation}
\label{60}
J_\pm(\eta)={1\over{2\pi}}(\partial_\pm\eta +\, e\,A_\pm)\;\;\; ; \;\;\eta=
\varphi , \rho
\end{equation}

\noindent  The important step now is to introduce the 
soldering field $B_\pm$ coupled with the currents so that,

\begin{equation}
\label{70}
W_\pm^{(1)}[\eta] = W_\pm[\eta] -\int d^2x\, B_\mp\, J_\pm(\eta)
\end{equation}

\noindent Then it is possible to define a modified action,

\begin{equation}
\label{80}
W[\varphi,\rho]= W_+^{(1)}[\varphi] + W_-^{(1)}[\rho]
 + {1\over{2\pi}} \int d^2x \, B_+ \,B_-
\end{equation}

\noindent which is invariant under an extended set of transformations that 
includes (\ref{40}) together with,

\begin{equation}
\delta B_{\pm}= \partial_{\pm}\alpha
\label{90}
\end{equation}

\noindent  Observe that the soldering field transforms exactly as a potential.
It has served its purpose of fusing the two chiral components. Since it 
is an auxiliary field, it can be eliminated from the invariant action 
(\ref{80}) by using the equations of motion. This will naturally solder the
otherwise independent chiral components and justifies its name as a soldering
field. The relevant solution is found to be,

\begin{equation}
\label{100}
B_\pm= 2\pi J_\pm
\end{equation}

\noindent Inserting this solution in (\ref{80}), we obtain,

\begin{equation}
\label{110}
W[\Phi]={1\over {4\pi}}\int d^2x\:\Big{\{}\Big{(}\partial_+
\Phi\partial_-\Phi + 2\,e\, A_+\partial_-\Phi - 2\,e\, A_-
\partial_+\Phi\Big{)} +(a+b-2)\,e^2\,A_+\,A_-\Big{\}}
\end{equation}

\noindent where,

\begin{equation}
\label{120}
\Phi=\varphi - \rho
\end{equation}

\noindent As announced, the action is no longer expreessed in terms of the
different scalars $\varphi$ and $\rho$, but only on their specific combination. 
This combination is gauge invariant. 

Let us digress on the significance of the findings. At the classical fermionic
version, the chiral lagrangeans are completely independent. Bosonising them
includes quantum effects, but still there is no correlation. The soldering
mechanism exploits the symmetries of the independent actions to
precisely combine them to yield a single action.
Note that the soldering works with the bosonised expressions. Thus the soldered
action obtained in this fashion corresponds to the quantum theory.    

We now show that different choices for the parameters $a$ and $b$
 lead to well known models.  To do this consider the variation of
(\ref{110}) under the conventional gauge transformations,
$\delta\varphi=\delta\rho=\alpha$ and $\delta A_\pm = \partial_\pm\alpha$.
It is easy to see that the expression
in parenthesis is gauge invariant. Consequently
a gauge invariant
structure for $W$ is obtained provided,
\begin{equation}
\label{130}
a+b-2=0
\end{equation}

The effect of soldering, therefore, has been to induce a lift of the initial
global symmetry (\ref{40}) to its local form. By functionally integrating
out the $\Phi$ field from (\ref{110}), we obtain,
\begin{equation}
\label{140}
W[A_+,A_-]=  -{ e^2\over 4\pi} \int d^2x\: \{A_+ 
{\partial_-\over \partial_+}A_+ 
+ A_- {\partial_+\over \partial_-}A_- - 2 A_+ A_-\}
\end{equation} 
which is the familiar structure for the gauge invariant action expressed in
terms of the potentials. The opposite chiralities of the 
two independent fermionic theories have been soldered to yield a gauge 
invariant action.

Some interesting observations are possible concerning the regularisation
ambiguity manifested by the parameters $a$ and $b$. Since a single equation
(\ref{130}) cannot fix both the parameters, it might appear that there 
is a whole one parameter class of
solutions for the chiral actions that combine to 
yield the vector gauge invariant
action.  Indeed, without any further input, this is the only conclusion.
However, Bose symmetry imposes a crucial restriction. This symmetry plays
an essential part that complements gauge invariance. Recall, for instance,
the calculation of the triangle graph leading to the Adler-Bell-Jackiw
anomaly. The familiar form of the anomaly cannot be obtained by simply
demanding gauge invariance; Bose symmetry at the vertices of the triangle
must also be imposed\cite{LR,SA}. Similarly, Bose symmetry\cite{RB3}
is essential in reproducing
the structure of the one-cocycle that is mandatory in the analysis on
smooth bosonisation\cite{DNS}; gauge invariance alone fails. In the present case,
this symmetry corresponds to the left-right (or + -) symmetry in (\ref{30}),
thereby requiring $a=b$. Together with the condition (\ref{130}) this implies
$a=b=1$. This parametrisation has important consequences if a Maxwell term
was included from the beginning to impart dynamics. Then the soldering takes
place among two chiral Schwinger models\cite{JR} having opposite chiralities to
reproduce the usual Schwinger model\cite{JS}. It
is known that the chiral models satisfy unitarity provided $a, b \geq 1$ and
the spectrum consists of a vector boson with mass,
\be
m^2 = \frac{e^2 a^2}{a-1}
\label{150}
\ee
and a massless chiral boson.  The values of the parameters obtained here
just saturate the bound. 
In other words, the chiral Schwinger model may have any 
$a\geq 1$, but if two such models with opposite chiralities are soldered to 
yield the vector Schwinger model, then the minimal bound is the unique
choice.
Moreover, for the minimal parametrisation,
the mass of the vector boson 
becomes infinite so that it goes out of the spectrum. Thus the
soldering mechanism shows how the massless modes in the chiral Schwinger
models are fused to generate the massive mode of the Schwinger model.

Naively it may appear that the soldering of the left and right chiralities
to obtain a gauge invariant result is a
simple issue since adding the classical lagrangeans
$\bar\psi\Dslash_+\psi$ and $\bar\psi\Dslash_-\psi$, with identical
fermion species, just yields the
usual vector lagrangean $\bar\psi\Dslash\psi$. The quantum considerations 
are, however, much involved. The chiral determinants, as they occur,
cannot be even defined
since the kernels map from one chirality to the other so that there is no
well defined eigenvalue problem\cite{AW,RB3}. This is circumvented by
working with
$\bar\psi(i\dslash + e\aslash_{\pm})\psi$, that satisfy an eigenvalue
equation, from which their determinants may be computed. But now a simple
addition of the classical lagrangeans does not reproduce the expected
gauge invariant form. At this juncture, the soldering process becomes
important. It systematically combined the quantised (bosonised)
expressions for the opposite chiral components. Note that {\it different}
fermionic species were considered so that this soldering does not have
any classical analogue, and is strictly a quantum phenomenon. This will 
become more transparent when the three dimensional case is discussed.

It is interesting to show that a different choice for the parameters $a$
and $b$ in (\ref{110}) leads to the Thirring model. Indeed it is precisely
when the mass term exists ($i.e., \,\, a+b-2\neq 0$), that (\ref{110})
represents
the Thirring  model. Consequently, this parametrisation complements that used
previously to obtain the vector gauge invariant structure. It is now easy to
see that the term in parentheses in (\ref{110}) corresponds to $\bar\psi
(i\dslash +e\aslash) \psi$ so that integrating out the auxiliary $A_\mu$ field
yields,
\be
{\cal L}=\bar\psi i\dslash\psi - \frac{g}{2}(\bar\psi\gamma_\mu\psi)^2\,\,\,\,
;\, g=\frac{4\pi}{a+b-2}
\label{A1}
\ee
which is just the lagrangean for the usual Thirring model. It is known
\cite{SC}that 
this model is meaningful provided the coupling parameter satisfies the 
condition $g>-\pi$, so that,
\be
\label{A2}
\mid a+b \mid >2
\ee
This condition is the analogue of (\ref{130}) found earlier. As usual, there
is a one parameter arbitrariness. Imposing Bose symmetry implies that both
$a$ and $b$ are equal and lie in the range
\be
\label{A3}
1<\mid a\mid =\mid b\mid
\ee
This may be compared with the previous case where $a=b=1$ was necessary for
getting the gauge invariant structure.  Interestingly, the positive range
for the parameters in (\ref{A3}) just commences from this value.

Having developed and exploited the concepts of soldering in two dimensions,
it is natural to investigate their consequences in three dimensions. The
discerning reader may have noticed that it is essential to have dual
aspects of a symmetry that can be soldered to yield new information. In the
two dimensional case, this was the left and right chirality. Interestingly,
in three dimensions also, we have a similar phenomenon.

\bigskip
\bigskip

\section{The three dimensional example}

\bigskip

This section is devoted to an analysis of the soldering process in the massive
Thirring model in three dimensions.  We shall show that two apparently
independent massive Thirring models in the long wavelength limit combine,
at the quantum level, into a massive Maxwell theory. 
This is further vindicated by a direct comparison of the current correlation
functions following from the bosonization identities. 
These findings are also extended to include three dimensional
quantum electrodynamics. 
The new results and interpretations illuminate a close
parallel with the two dimensional discussion.

\bigskip

\subsection{The massive Thirring model}

\bigskip

In order to effect the soldering, the first step is to consider the
bosonisation of the massive Thirring
model in three dimensions\cite{C,RB}. This is therefore reviewed briefly. The
relevant current correlator generating functional,
in the Minkowski metric, is given by,
\be
\label{160}
Z[\kappa]=\int D\psi D\bar\psi \exp\Bigg(i\int d^3x\Bigg
[\bar\psi(i \dslash + m )\psi -\frac{\lambda^2}{2}
j_\mu j^\mu + \lambda j_\mu \kappa^\mu\Bigg]\Bigg)
\ee
where $j_\mu=\bar\psi\gamma_\mu\psi$ is the fermionic current. As usual,
the four fermion interaction can be eliminated by introducing an auxiliary
field,
\be
\label{170}
Z[\kappa]=\int D\psi D\bar\psi Df_\mu\exp\Bigg(i\int d^3x\Bigg
[\bar\psi\left(i \dslash + m +\lambda (\fslash +\kslash)\right)
\psi +\frac{1}{2} f_\mu f^\mu\Bigg]\Bigg)
\ee
Contrary to the two dimensional models, the fermion integration cannot be
done exactly. Under certain limiting conditions, however, this integation
is possible leading to closed expressions. A particularly effective choice
is the large mass limit in which case the fermion determinant yields a local
form. Incidentally, any other value of the mass leads to a nonlocal structure
\cite{RB1}.
The large mass limit is therefore very special. The leading term in this
limit was calculated by various means \cite{DJT} 
and shown to yield the Chern-Simons
three form. Thus the generating functional for the massive Thirring model in
the large mass limit is given by,
\be 
\label{180}
Z[\kappa]=\int Df_\mu 
\exp\Bigg( i\int d^3x\:\Bigg({\lambda^2\over{8\pi}}{m\over{\mid m\mid}}
\epsilon_{\mu\nu\lambda}f^\mu\partial^\nu f^\lambda +
\frac{1}{2} f_\mu f^\mu +\frac{\lambda^2}{4\pi}\frac{m}{\mid m\mid}
\epsilon_{\mu\nu\sigma}\kappa^\mu\partial^\nu f^\sigma\Bigg)\Bigg)
\ee
where the signature of the topological terms is dictated by the corresponding
signature of the fermionic mass term. 
In obtaining the above result a local counter term has been ignored. 
Such terms manifest the ambiguity in defining the time ordered product
to compute the correlation functions\cite{BRR}.
The lagrangean in the above partition
function defines a self dual model introduced earlier \cite{TPN}. The massive
Thirring model, in the relevant limit, therefore bosonises to a self dual
model. It is useful to clarify the meaning of this self duality. The 
equation of motion in the absence of sources is given by,
\be
\label{190}
f_\mu =-{\lambda^2\over{4\pi}}{m\over{\mid m\mid}}
\epsilon_{\mu\nu\lambda}\partial^\nu f^\lambda  
\ee
from which the following relations may be easily verified,
\br
\label{200}
\partial_\mu f^\mu &=& 0\nonumber\\
\left(\Box + M^2\right)f_\mu &=& 0 \,\,\,\,\,\, ;\,\, 
M=\frac{4\pi}{\lambda^2}
\er
A field dual to $f_\mu$ is defined as,
\be
\label{210}
\tilde f_\mu = {1\over M} \epsilon_{\mu\nu\lambda}\partial^\nu f^\lambda
\ee
where the mass parameter $M$ is inserted for dimensional reasons. Repeating
the dual operation, we find,
\be
\label{220}
\tilde{\left(\tilde{f_\mu}\right)}= 
{1\over M} \epsilon_{\mu\nu\lambda}\partial^\nu\tilde{f^\lambda}=f_\mu
\ee
obtained by exploiting (\ref{200}), thereby validating the definition of
the dual field.  Combining these results with (\ref{190}),
we conclude that,
\be
\label{230}
f_\mu=- \frac{m}{\mid m \mid} \tilde f_\mu
\ee
Hence, depending on the sign of the fermion mass term, the bosonic theory
corresponds to a self-dual or an anti self-dual model.  Likewise, the Thirring
current bosonises to the topological current

\be
\label{235}
j_\mu = \frac{\lambda}{4\pi}\frac{m}{\mid m\mid}\epsilon_{\mu\nu\rho}
\partial^\nu f^\rho
\ee

The close connection with the two dimensional analysis is now evident.
There the starting point was to consider two distinct fermionic theories with 
opposite chiralities. In the present instance, the analogous thing is to
take two independent Thirring models with identical coupling strengths but
opposite mass signatures,
\br
\label{240}
{\cal L_+}&=&\bar\psi\left(i\dslash + m\right)\psi -\frac{\lambda^2}{2}\left(\bar\psi\gamma_\mu\psi\right)^2\nonumber\\
{\cal L_-} &=& \bar \xi\left(i\dslash - m'\right)\xi - \frac{\lambda^2}{2} \left(\bar\xi\gamma_\mu\xi\right)^2
\er
Note that the only the relative sign between the mass parameters is important,
but their magnitudes are different. From now on it is also assumed that
both $m$ and $m'$ are positive.
Then the bosonised lagrangeans are, respectively,
\br
\label{250}
{\cal L_+}&=&\frac{1}{2M} 
\epsilon_{\mu\nu\lambda}f^\mu\partial^\nu f^\lambda +
{1\over 2} f_\mu f^\mu\nonumber\\
{\cal L_-} &=&- \frac{1}{2M}
\epsilon_{\mu\nu\lambda}g^\mu\partial^\nu g^\lambda +
{1\over 2} g_\mu g^\mu
\er
where $f_\mu$ and $g_\mu$ are the distinct bosonic vector fields.  The current
bosonization formulae in the two cases are given by

\br
\label{255}
j_\mu^+ &=& \bar\psi\gamma_\mu\psi=\frac{\lambda}{4\pi} 
\epsilon_{\mu\nu\rho}\partial^\nu f^\rho\nonumber\\
j_\mu^- &=&\bar\xi\gamma_\mu\xi= - \frac{\lambda}{4\pi} 
\epsilon_{\mu\nu\rho}\partial^\nu g^\rho
\er

The stage is now set for soldering. Taking a cue from the two dimensional
analysis, let us consider the gauging of the following symmetry,
\be
\label{260}
\delta f_\mu = \delta g_\mu = 
\epsilon_{\mu\rho\sigma}\partial^\rho \alpha^\sigma
\ee
Under such transformations, the bosonised lagrangeans change as,
\be
\label{270}
\delta{\cal L_\pm} = J_\pm^{\rho\sigma}(h_\mu)
 \partial_\rho\alpha_\sigma \,\,\,\,\, ;\,\, h_\mu=f_\mu,\,\,g_\mu
\ee
where the antisymmetric currents are defined by,
\be
\label{280}
J_\pm^{\rho\sigma}(h_\mu)= \epsilon^{\mu\rho\sigma}h_\mu \pm {1\over M}
\epsilon^{\gamma\rho\sigma}\epsilon_{\mu\nu\gamma}\partial^\mu 
h^\nu
\ee
It is worthwhile to mention that any other variation of the fields
(like $\delta{f_\mu}=\alpha_\mu$)is inappropriate because changes in
the two terms of the lagrangeans cannot be combined to give a single
structure like (\ref{280}). We now introduce the soldering field coupled
with the antisymmetric currents. In
the two dimensional case this was a vector. Its natural extension now
is the antisymmetric second rank Kalb-Ramond tensor field $B_{\rho\sigma}$,
transforming in the usual way,
\be
\label{290}
\delta B_{\rho\sigma}=\partial_\rho\alpha_\sigma -
\partial_\sigma\alpha_\rho
\ee
Then it is easy to see that the modified lagrangeans,
\be
\label{300}
{\cal L}_\pm^{(1)}={\cal L}_\pm - {1\over 2} J_\pm^{\rho\sigma}(h_\mu)
B_{\rho\sigma}
\ee
transform as,
\be
\label{310}
\delta{\cal L}_\pm^{(1)}=- {1\over 2} \delta J_\pm^{\rho\sigma}
B_{\rho\sigma}
\ee
The final modification consists in adding a term to ensure gauge invariance
of the soldered lagrangean. This is achieved by,
\be
\label{320}
{\cal L}_\pm^{(2)}={\cal L}_\pm^{(1)} + {1\over 4} 
B^{\rho\sigma}B_{\rho\sigma}
\ee
A straightforward algebra shows that the following combination,
\br
\label{330}
{\cal L}_S &=& {\cal L}_+^{(2)}+{\cal L}_-^{(2)}\nonumber\\
&=&{\cal L}_+ + {\cal L}_- -{1\over 2}B^{\rho\sigma}
\left(J^+_{\rho\sigma}(f) + J^-_{\rho\sigma}(g)\right)
+{1\over 2} B^{\rho\sigma}B_{\rho\sigma}
\er
is invariant under the gauge transformations (\ref{260}) and (\ref{290}). 
The gauging of the symmetry
is therefore complete. To return to a description in terms of the original
variables, the auxiliary soldering field is eliminated from (\ref{330}) by 
using the equation of motion,
\be
\label{340}
B_{\rho\sigma}= {1\over 2} \left(J_{\rho\sigma}^+(f)+
J_{\rho\sigma}^-(g)\right)
\ee
Inserting this solution in (\ref{330}), the final soldered
lagrangean is expressed
solely in terms of the currents involving the original fields,
\be
\label{350}
{\cal L}_S ={\cal L}_+ + {\cal L}_- -
{1\over 8}\left(J_{\rho\sigma}^+(f)+
J_{\rho\sigma}^-(g)\right)\left(J^{\rho\sigma}_+(f)+
J^{\rho\sigma}_-(g)\right)
\ee
It is now crucial to note that, by using the explicit structures for the
currents, the above lagrangean is no longer a function of $f_\mu$ and $g_\mu$
separately, but only on the combination,
\be
\label{360}
A_\mu = {1\over{\sqrt{2} M}}\left(g_\mu - f_\mu\right)
\ee
with,
\be
\label{370}
{\cal L}_S = - \frac{1}{4} F_{\mu\nu}F^{\mu\nu} + 
{M^2\over 2}A_\mu A^\mu
\ee
where,
\be
\label{380}
F_{\mu\nu}= \partial_\mu A_\nu -\partial_\nu A_\mu
\ee
is the usual field tensor expressed in terms of the basic entity $A_\mu$.
Our goal has been achieved. The soldering mechanism has precisely fused
the self and anti self dual symmetries to yield a massive Maxwell theory
which, naturally, lacks this symmetry.

It is now instructive to understand this result by comparing the current
correlation functions.  The Thirring currents in the two models bosonise
to the topological currents (\ref{255}) in the dual formulation.  From a
knowledge of the field correlators in the latter case, it is therefore
possible to obtain the Thirring current correlators.  The field
correlators are obtained from the inverse of the kernels occurring in
(\ref{250}),

\br
\label{381}
\langle f_\mu(+k)\: f_\nu(-k)\rangle &=&
\frac{M^2}{M^2 - k^2}\left(i g_{\mu\nu} +
{1\over M}\epsilon_{\mu\rho\nu} k^\rho -
\frac{i}{M^2} k_\mu\:k_\nu\right)\nonumber\\
\langle g_\mu(+k)\: g_\nu(-k)\rangle &=&
\frac{M^2}{M^2 - k^2}\left(i g_{\mu\nu} -
{1\over M}\epsilon_{\mu\rho\nu} k^\rho -
\frac{i}{M^2} k_\mu\:k_\nu\right)
\er

\noindent where the expressions are given in the momentum space.
Using these in (\ref{255}), the current
correlators are obtained, 

\br
\label{382}
\langle j_\mu^+(+k) j_\nu^+(-k)\rangle &=&
\frac{M}{4\pi (M^2 - k^2)}
\left(i k^2 g_{\mu\nu}- ik_\mu\: k_\nu
+{1\over M}\epsilon_{\mu\nu\rho}
k^\rho\:k^2\right)\nonumber\\
\langle j_\mu^-(+k) j_\nu^-(-k)\rangle &=&
\frac{M}{4\pi (M^2 - k^2)}
\left(i k^2 g_{\mu\nu}- ik_\mu\: k_\nu
-{1\over M}\epsilon_{\mu\nu\rho}
k^\rho\:k^2\right)
\er
It is now feasible to construct a total current,

\be
\label{383}
j_\mu=j_\mu^+ + j_\mu^- = \frac{\lambda}{4\pi}
\epsilon_{\mu\nu\rho}\partial^\nu 
\left(f^\rho - g^\rho\right)
\ee
Then the correlation function for this current, in the original self dual
formulation, follows
from (\ref{382}) and noting that $\langle j_\mu^+\:
j_\nu^-\rangle =0$, which is a consequence of the
independence of $f_\mu$ and $g_\nu$;

\be
\label{384}
\langle j_\mu(+k)\: j_\nu(-k)\rangle = \langle j_\mu^+\: j_\nu^+\rangle +
\langle j_\mu^-\: j_\nu^-\rangle =
\frac{iM}{2\pi(M^2 -k^2)}
\left(k^2\: g_{\mu\nu} - k_\mu\: k_\nu\right)
\ee
The above equation is easily reproduced from the 
effective theory.  Using (\ref{360}), it is observed that the
bosonization of the composite current (\ref{383}) is
defined in terms of the massive vector field $A_\mu$,

\be
\label{385}
j_\mu=\bar\psi\gamma_\mu\psi +\bar\xi\gamma_\mu\xi =
-\sqrt{{M\over 2\pi}}\epsilon_{\mu\nu\rho}
\partial^\nu A^\rho
\ee
The current correlator is now obtained from the field
correlator $\langle A_\mu\: A_\nu\rangle$ given by the
inverse of the kernel appearing in (\ref{370}),

\be
\label{386}
\langle A_\mu(+k)\: A_\nu(-k)\rangle =
\frac{i}{M^2 - k^2}\left(g_{\mu\nu} - 
\frac{k_\mu\: k_\nu}{M^2}\right)
\ee
>From (\ref{385}) and (\ref{386}) the two point function
(\ref{384}) is reproduced, including the normalization.

We conclude, therefore, that two massive Thirring models with opposite 
mass signatures, in the long wavelength limit,
combine by the process of bosonisation and soldering, to a massive
Maxwell theory. The bosonization of the composite current, obtained
by adding the separate contributions from the two models, is given in
terms of a topological current(\ref{385}) of the massive vector theory.
These are completely new results which cannot be obtained by a
straightforward application of conventional bosonisation techniques.
The massive modes in the original Thirring models are 
manifested in the two modes of (\ref{370}) so that there is a proper
matching in the degrees of freedom.
Once again it is reminded that the 
fermion fields for the models are different so that the analysis has no
classical analogue. Indeed if one considered the same fermion species,
then a simple addition of the classical lagrangeans would lead to a 
Thirring model with a mass given by $m-m'$. 
In particular, this difference can be zero.
The bosonised version of such a massless model is
known \cite{M, RB1} to yield a highly nonlocal theory which has no connection
with (\ref{370}). Classically, therefore, there is no possibility of even
understanding, much less, reproducing the effective quantum result. 
In this sense the application in three dimensions
is more dramatic than the corresponding case of two dimensions.
        
\bigskip

\subsection{Quantum electrodynamics}

\bigskip

An interesting theory in which the preceding ideas may be
implemented is quantum electrodynmics, whose current
correlator generating
functional in an arbitrary covariant gauge is given by,
\be
\label{390}
Z[\kappa]=\int D\bar\psi\: D\psi\: DA_\mu\, \exp
\left\{i\int d^3x\:\left(\bar\psi\left(i\dslash + m +e\aslash
\right)\psi -{1\over 4} F_{\mu\nu}F^{\mu\nu} 
+{\eta\over 2}(\partial_\mu A^\mu)^2 + e j_\mu\kappa^\mu
\right)\right\}
\ee
where $\eta$ is the gauge fixing parameter and
$j_\mu = \bar\psi\gamma_\mu\psi$ is the current. 
As before, a one loop computation of the fermionic
determinant in the large mass limit yields,
\br
\label{400}
Z[\kappa]&=&\int  DA_\mu\, 
\exp { \lbrace} i\int d^3x\:\lbrack\frac{e^2}{8\pi}
\frac{m}{\mid m\mid}
\epsilon_{\mu\nu\lambda}A^\mu\partial^\nu A^\lambda
-{1\over 4} F_{\mu\nu}F^{\mu\nu}\nonumber\\
&+& \frac{e^2}{4\pi}
\frac{m}{\mid m \mid} \epsilon_{\mu\nu\rho}
\kappa^\mu\partial^\nu\: A^\rho +{\eta\over 2}
(\partial_\mu A^\mu)^2\rbrack{\rbrace}
\er
In the absence of sources, this just corresponds to the
topolologically massive Maxwell-Chern-Simons theory, with
the signature of the topological term determined from that of
the fermion mass term. The equation of motion,

\be
\label{405}
\partial^\nu\, F_{\nu\mu} +\frac{e^2}{4\pi}
\frac{m}{\mid m \mid} \epsilon_{\mu\nu\lambda}
\partial^\nu A^\lambda = 0
\ee
expressed in terms of the dual tensor,

\be
\label{410}
F_\mu = \epsilon_{\mu\nu\lambda}
\partial^\nu A^\lambda
\ee
reveals the self (or anti self) dual property,
\be
\label{420}
F_\mu =\frac{4\pi}{e^2}\frac{m}{\mid m\mid}
\epsilon_{\mu\nu\lambda}\partial^\nu F^\lambda
\ee
which is the analogue of (\ref{190}).  In this fashion the Maxwell-Chern-Simons theory
manifests the well known \cite{DJ, BRR, BR} mapping with the self
dual models considered in the previous subsection.
The difference is that the self duality in the former,
in contrast to the latter, is contained in the dual
field (\ref{410}) rather than in the basic field defining
the theory. This requires some modifications in the
ensuing analysis.  Furthermore, the bosonization of the
fermionic current is now given by the topological current
in the Maxwell-Chern-Simons theory,

\be
\label{425}
j_\mu = \frac{e}{4\pi} \frac{m}{\mid m \mid}
\epsilon_{\mu\nu\lambda}\partial^\nu A^\lambda
\ee

Consider, therefore, two independent models describing quantum 
electrodynamics with opposite signatures in the mass terms,

\br
\label{426}
{\cal L}_+ &=& \bar\psi\left( i\dslash +m
+e\aslash\right)\psi -{1\over 4} 
F_{\mu\nu}(A) F^{\mu\nu}(A)\nonumber\\
{\cal L}_- &=& \bar\xi\left( i\dslash -m'
+e\bslash\right)\xi -{1\over 4} 
F_{\mu\nu}(B) F^{\mu\nu}(B)
\er

whose bosonised versions in an appropriate limit are given by,
\br
\label{430}
{\cal L}_+ &=& -\frac{1}{4} F_{\mu\nu}(A)+\frac{M}{2}
\epsilon_{\mu\nu\lambda}A^\mu\partial^\nu A^\lambda 
\,\,\,\,\, ; \,\, M=\frac{e^2}{4\pi}\nonumber\\
{\cal L}_- &=& -\frac{1}{4} F_{\mu\nu}(B)-\frac{M}{2}
\epsilon_{\mu\nu\lambda}B^\mu\partial^\nu B^\lambda
\er
where $A_\mu$ and $B_\mu$ are the corresponding potentials. 
Likewise, the corresponding expressions for the bosonized
currents are found from the general structure (\ref{425}),

\br
\label{435}
j_\mu^+ &=&\bar\psi\gamma_\mu\psi= \frac{M}{e} \epsilon_{\mu\nu\lambda}
\partial^\nu A^\lambda\nonumber\\
j_\mu^- &=&\bar\xi\gamma_\mu\xi= -\frac{M}{e} \epsilon_{\mu\nu\lambda}
\partial^\nu B^\lambda
\er
To proceed with
the soldering of the above models, take  the symmetry transformation,
\be
\label{440}
\delta A_\mu=\alpha_\mu
\ee
Such a transformation is spelled out by recalling (\ref{260}) and the 
observation that now (\ref{410}) simulates the $f_\mu$ field in the previous
case. Under this variation, the lagrangeans (\ref{430}) change as,
\be
\label{450}
\delta{\cal L}_\pm =J_\pm^{\rho\sigma}(P)\partial_\rho\alpha_\sigma
\,\,\,\,\, ; \,\, P=A,B
\ee
where the antisymmetric  currents are defined by,
\be
\label{460}
J_\pm^{\rho\sigma}(P)=\pm m \epsilon^{\rho\sigma\mu}P_\mu -F^{\rho\sigma}(P)
\ee
Proceeding as before, the antisymmetric soldering field $B_{\alpha\beta}$
transforming as (\ref{290}) is introduced by coupling 
with these currents to define the
first iterated lagrangeans analogous to (\ref{300}),
\be
\label{470}
{\cal L}_\pm^{(1)}={\cal L}_\pm - 
{1\over 2} J_\pm^{\rho\sigma}(P) B_{\rho\sigma}
\ee
These lagrangeans are found to transform as,
\be
\label{480}
\delta{\cal L}_\pm^{(1)} ={1\over 4} \delta B^2_{\lambda\sigma}
-{1\over 2}\left(\pm m \epsilon_{\mu\lambda\sigma}\alpha^\mu
B^{\lambda\sigma}\right)
\ee
It is now straightforward to deduce the final lagrangean that will be
gauge invariant. This is given by,
\be
\label{490}
{\cal L}_S = {\cal L}_+^{(2)} + {\cal L}_-^{(2)}
\,\,\,\,\, ; \,\, \delta{\cal L}_S =0
\ee
where the second iterated pieces are,
\be
\label{500}
{\cal L}_\pm^{(2)}= {\cal L}_\pm -{1\over 2} J_\pm^{\rho\sigma}
B_{\rho\sigma} -{1\over 4}B_{\rho\sigma}B^{\rho\sigma}
\ee
The invariance of ${\cal L}_S$ (\ref{490}) is verified by observing that,
\be
\label{510}
\delta{\cal L}_\pm^{(2)}=\mp{1\over 2}m\epsilon_{\mu\lambda\sigma}
\alpha^\mu B^{\lambda\sigma}
\ee
To obtain the effective lagrangean it is necessary to eliminate the auxiliary
$B_{\rho\sigma}$ field by using the equation of motion following from
(\ref{490}),
\be
\label{520}
B_{\sigma\lambda}=-{1\over 2}\left(J^+_{\sigma\lambda}(A) +
J^-_{\sigma\lambda}(B)\right)
\ee
Putting this back in (\ref{490}), we obtain the final soldered lagrangean,
\be
\label{530}
{\cal L}_S =-{1\over 4} F_{\mu\nu}(G)F^{\mu\nu}(G) 
+\frac{M^2}{2}G_\mu G^\mu
\ee
written in terms of a single field,
\be
\label{540}
G_\mu =\frac{1}{\sqrt{2}}\left(A_\mu - B_\mu\right)
\ee
The lagrangean (\ref{530}) governs the dynamics of a massive Maxwell theory.

As before, we now discuss the implications for the current
correlation functions.  These functions in the original models
describing electrodynamics can be obtained from the mapping
(\ref{435}).  The first step is to abstract the basic field
correlators found by inverting the kernels occurring in
(\ref{430}).  The results, in the momentum space, are

\br
\label{545}
\langle A_\mu(+k)\: A_\nu(-k)\rangle &=&
\frac{i}{M^2 -k^2}\left[ g_{\mu\nu} +
\frac{M^2 -k^2(\eta +1)}{\eta k^4}k_\mu \: k_\nu +
\frac{iM}{k^2} \epsilon_{\mu\rho\nu}k^\rho\right]
\nonumber\\
\langle B_\mu(+k)\: B_\nu(-k)\rangle &=&
\frac{i}{M^2 -k^2}\left[ g_{\mu\nu} +
\frac{M^2 -k^2(\eta +1)}{\eta k^4}k_\mu \: k_\nu -
\frac{iM}{k^2} \epsilon_{\mu\rho\nu}k^\rho\right]
\er
The current correlators are easily computed by
substituting (\ref{545}) into (\ref{435}),

\br
\label{546}
\langle j_\mu^+(+k)\:j_\nu^+(-k)\rangle &=&
i\left(\frac{M}{e}\right)^2\frac{1}{M^2 -k^2}
\left[k^2 g_{\mu\nu} -k_\mu\: k_\nu -
i M\epsilon_{\mu\nu\rho}k^\rho\right]
\nonumber\\
\langle j_\mu^-(+k)\:j_\nu^-(-k)\rangle &=&
i\left(\frac{M}{e}\right)^2\frac{1}{M^2 -k^2}
\left[k^2 g_{\mu\nu} -k_\mu\: k_\nu +
i M\epsilon_{\mu\nu\rho}k^\rho\right]
\er
where, expectedly, the gauge dependent ($\eta$) contribution has
dropped out.  Defining a composite current,

\be
\label{547}
j_\mu = j_\mu^+ + j_\mu^- = \frac{M}{e} 
\epsilon_{\mu\nu\lambda}\partial^\nu
\left(A^\lambda - B^\lambda\right)
\ee
it is simple to obtain the relevant correlator
by exploiting the results for $j_\mu^+$ and $j_\mu^-$ from (\ref{546}),

\be
\label{548}
\langle j_\mu(+k)\: j_\nu(-k)\rangle =
2i\left(\frac{M}{e}\right)^2\frac{1}{M^2 -k^2} 
\left(k^2 g_{\mu\nu} - k_\mu\: k_\nu\right)
\ee
In the bosonized version obtained from the soldering,
(\ref{547}) represents the mapping,

\be
\label{549}
j_\mu=\bar\psi\gamma_\mu\psi +\bar\xi\gamma_\mu\xi =
\sqrt 2 {M\over e} \epsilon_{\mu\nu\lambda}
\partial^\nu G^\lambda
\ee
where $G_\mu$ is the massive vector field (\ref{540})
whose dynamics is governed by the lagrangean (\ref{530}). 
In this effective description the result (\ref{548}) is
reproduced from (\ref{549}) by using the correlator
of $G_\mu$ obtained from (\ref{530}), which is exactly identical to
(\ref{386}).

Thus the combined effects of bosonisation and soldering show that two
independent
quantum electrodynamical models with appropriate mass signatures are
equivalently described by the massive Maxwell theory. In the self dual version
the massive modes are the topological excitations in the Maxwell-Chern-Simons
theories. These are combined into the two usual massive modes in the effective
massive vector theory.
A complete correspondence among the composite current correlation
functions in the original models and in their dual bosonised
description was also established.  The comments made
in the concluding part of the last subsection naturally apply also in this
instance.

\bigskip

\section{Conclusions}

\bigskip

The present analysis clearly revealed the possibility of obtaining new
results from quantum effects that conspire to combine two apparently
independent theories into a single effective theory. 
The essential ingredient was that these theories
must possess the dual aspects of the same symmetry. Then, by a systematic
application of bosonisation and soldering, it was feasible to abstract
a meaningful combination of such models, which can never be obtained
by a naive addition of the classical lagrangeans.

The basic notions and ideas were particularly well illustrated in the
two dimensional example where the bosonised expressions for distinct
chiral lagrangeans were soldered to reproduce either the usual gauge
invariant theory or the Thirring model. Indeed, the soldering mechanism
that fused the opposite chiralities, clarified several aspects of the  
ambiguities occurring in bosonising chiral lagrangeans. It was clearly 
shown that unless Bose symmetry is imposed as an additional restriction,
there is a whole one parameter class of bosonised solutions for the chiral
lagrangeans that can be soldered to yield the vector gauge invariant result.
The close connection between Bose symmetry and gauge invariance was thereby
established, leading to a unique parametrisation. Similarly, using a 
different parametrisation, the soldering of the chiral lagrangeans led to
the Thirring model. Once again there was a one parameter ambiguity unless
Bose symmetry was imposed. If that was done, there was a specified range of 
solutions for the chiral lagrangeans that combined to yield a well defined
Thirring model.

The elaboration of our methods was done by considering the massive version
of the Thirring model and quantum electrodynamics in three dimensions.
By the process of bosonisation such models, in the long wavelength limit, 
were cast in a form which 
manifested a self dual symmetry. This was a basic perquisite for effecting
the soldering. It was explicitly shown that two distinct massive Thirring 
models, with opposite mass signatures, combined to a massive Maxwell theory.
The Thirring current correlation functions calculated either in the original
self dual formulation or in the effective massive vector theory yielded 
identical results, showing the consistency of our approach. The application
to quantum electrodynamics followed along similar lines.

It is evident that the present technique of combining models by the two
step process of bosonisation and soldering can be carried through in
higher dimensions provided the models have the relevant symmetry properties.
It is also crucial to note that duality pervades the entire analysis.
In the three dimensional case this was self evident since the models
had a self (and  anti) self dual symmetry. This was hidden in the two
dimensional case where chiral symmetry was more transparent. But it may be
mentioned  that in two dimensions, chiral symmetry is the analogue of 
the duality $\partial_\mu\phi=\pm\epsilon_{\mu\nu}\partial^\nu\phi$. 
Interestingly, the duality
in two dimensions was manifest in the lagrangeans while that in three 
dimensions was contained in the equations of motion. This opens up
the possibility to discuss different aspects of duality, contained either
in the lagrangean or in the equations of motion, in the same framework.
Consequently, the methods developed here can be relevant and useful in
diferent contexts; particularly in the recent discussions on electromagnetic
duality or the study of chiral forms which exactly possess the type of self
dual symmetry considered in this paper. We will report on these and related
issues in a future work.


\newpage
\begin{thebibliography}{30}
\bibitem{AAR} For a review see, E. Abdalla, M.C.B. Abdalla and K.D. Rothe, 
{\it Nonperturbative
Methods in Two Dimensional Quantum Field Theory}, World Scientific,
Singapore, 1991.
\bibitem{M} E.C. Marino, Phys. Lett. B263 (1991) 63.
\bibitem{C} C. Burgees, C. L\"utken and F. Quevedo, Phys. Lett. B336 (1994)
18; E. Fradkin and F. Schaposnik, Phys. Lett. B338 (1994) 253; K. Ikegami,
K. Kondo and A. Nakamura, Prog. Theor. Phys. 95 (1996) 203; D.Barci, C.D.
Fosco and L. Oxman, Phys. Lett. B375 (1996) 267.
\bibitem{RB} R. Banerjee, Phys. Lett. B358 (1995) 297 and Nucl. Phys.
B465 (1996) 157.
\bibitem{RB1} R. Banerjee and E.C. Marino, hep-th/9607040 (To appear in
Phys. Rev. D); hep-th/9707033 and 9707100.
\bibitem{AB} E. Abdalla and R. Banerjee, hep-th/9704176, Phys. Rev. Lett.,
to appear.
\bibitem{S} M. Stone, University of Illinois Preprint, ILL-TH-28-89.
\bibitem{W} R. Amorim, A. Das and C. Wotzasek, Phys. Rev. D53 (1996) 5810.
\bibitem{RJ} See, for instance, R. Jackiw, {\it Diverse Topics in Theoretical
and Mathematical Physics}, World Scientific, Singapore, 1995.
\bibitem{RB2} R. Banerjee, Phys. Rev. Lett. 56 (1986) 1889.
\bibitem{ABW} E.M.C. Abreu, R. Banerjee and C. Wotzasek, hep-th/9707204,
Nucl.Phys. B, to appear.
\bibitem{LR} L. Rosenberg, Phys. Rev. 129 (1963) 2786.
\bibitem{SA} S.L. Adler, {\it Lectures in Elementary Particles and Quantum
Field Theory}, S.Deser et al. eds., 1970, Brandeis Lectures, MIT Press,
Cambridge.
\bibitem{RB3} N. Banerjee and R. Banerjee Nucl. Phys. B445 (1995) 516.
\bibitem{DNS} P.H.Damgaard, H.B.Nielsen, and R.Sollacher, 
Nucl. Phys. B414 (1994) 541;
P.H.Damgaard and R.Sollacher, Phys. Lett. B 322 (1994) 131
and Nucl. Phys. B 433 (1995) 671. 
\bibitem{JR} R. Jackiw and R. Rajaraman, Phys. Rev. Lett. 54 (1985) 1219;
2060(E).
\bibitem{JS}J.Schwinger, Phys. Rev. 128 (1962) 2425.
\bibitem{AW} L. Alvarez-Gaume and E. Witten, Nucl.Phys.B234 (1983) 269.
\bibitem{SC} S. Coleman, Phys. Rev. D11 (1975) 2088.
\bibitem{DJT} S. Deser, R. Jackiw and S. Templeton, Ann. Phys. (NY)
140 (1982) 372; A.N. Redlich, Phys. Rev. D29 (1984) 2366.
\bibitem{BRR} R. Banerjee, H.J. Rothe and K.D. Rothe, Phys. Rev. D52
(1995) 3750.
\bibitem{TPN} P.K. Townsend, K. Pilch and P. van Nieuwenhuizen, Phys.
Lett. B136 (1984) 452.
\bibitem{DJ} S. Deser and R. Jackiw, Phys. Lett. B139 (1984) 371.
\bibitem{BR} R. Banerjee and H.J. Rothe, Nucl. Phys. B447 (1995) 183.

\end{thebibliography}
\end{document}